\documentclass[twocolumn,nofootinbib,superscriptaddress,groupedaddress]{revtex4-1}

\pdfoutput=1

\usepackage[utf8]{inputenc}

\usepackage{slashed}

\usepackage[normalem]{ulem}

\usepackage{color}

\usepackage{epsfig}
\usepackage{epstopdf}
\usepackage{epsf}
\input{epsf.sty}
\usepackage{graphicx,amsmath,amssymb}
\usepackage{epsfig}
\allowdisplaybreaks

\def\ei{\end{itemize}}
\def\be{\begin{equation}}
\def\ee{\end{equation}}
\newcommand{\bea}{\begin{eqnarray}}
\newcommand{\eea}{\end{eqnarray}}

\usepackage{bbm,bm,amsmath,amssymb}

\usepackage{hyperref}

\usepackage{float}

\begin{document}

\title{Axion-Dilaton Destabilization and the Hubble Tension}
\author{Stephon Alexander}
\email{stephon\_alexander@brown.edu}

\author{Evan McDonough }
\email{evan\_mcdonough@brown.edu}

\affiliation{Brown Theoretical Physics Center, Brown University, Providence, RI, USA. 02912\\ }
\affiliation{ \\ Department of Physics, Brown University, Providence, RI, USA. 02912}

 \begin{abstract}
The discrepancy in measurements of the Hubble constant indicates new physics in dark energy, dark matter, or both. Drawing inspiration from string theory, where axions interact with the other moduli fields, including the dilaton, here we demonstrate that the dynamics of an interacting dilaton and axion naturally realizes the proposal of Early Dark Energy. In this setup, stabilization of the the dilaton is in part due to the axion, and in the early universe the dilaton contributes to dark energy. The combined axion-dilaton system is destabilized when the Hubble constant falls below the mass of the axion, triggering a phase of fast-roll evolution of the dilaton wherein its equation of state is $w=1$, and the early dark energy redshifts away as $a^{-6}$.
 \end{abstract}

\maketitle


\section{Introduction}
\label{sec:Introduction}

With the refined measurement of the distance to the Large Magellanic Cloud \cite{Riess:2019cxk}, the discrepancy between the observed and inferred values of the Hubble constant has raised in significance to 4.4$\sigma$. The data supporting this discrepancy comes from a wide range of redshift, providing the greatest challenge yet to the $\Lambda$CDM model. This growing observational evidence motivates the search for theoretical explanations.

In contrast with past cosmological glitches, e.g. anomalies in the Cosmic Microwave Background, there are still relatively few working theoretical explanations \cite{Poulin:2018cxd,DEramo:2018vss,Colgain:2018wgk,Kreisch:2019yzn,Vattis:2019efj,Agrawal:2019lmo,Kaloper:2019lpl}. However, some of the necessary ingredients of a solution are now known. As emphasized in \cite{Aylor:2018drw}, to rectify the cosmological distance ladder with the inverse distance ladder and the CMB requires a modification of the sound horizon $r_s$, and thus any solution must include new dynamics in the very early universe\footnote{We thank Matias Zaldarriaga and Christina D. Kreisch for emphasizing this to us.} \footnote{See also e.g. \cite{Mortsell:2018mfj}.}.

The scenario of `Early Dark Energy'  \cite{Karwal:2016vyq,Poulin:2018cxd} implemented a modification of $r_s$ via the decay of dark energy into a fluid that redshifts as fast or faster than radiation, i.e. with equation of state $w \geq 1/3$. This fluid can arise via a scalar field with potential that is a polynomial in the cosine of the field, $V = \mu^4 (\cos \phi)^p$ \cite{Poulin:2018cxd}, with parameters such that the decay happens shortly before recombination.

As an alternative to such a potential, \cite{Agrawal:2019lmo} demonstrated that single-field models with polynomial potentials can realize the early dark energy dynamics. The best fit model was found to be $V = {\rm eV}^4 (\phi/m_{pl})^4$, substantiated by a complete perturbation analysis, in which the perturbed field equations were solved directly.

In this letter we build on these previous works by demonstrating that the requisite dynamics can be realized via the interacting dynamical system of a dilaton field, referred to as such due to an exponential potential $V= V_0 e^{- \lambda \phi/m_{pl}}$, and an axion field with cosine potential. At early stages the axion is held up its potential by Hubble friction, and due to an interaction,  the dilaton is also held up its potential, and thus acts as dark energy. Eventually, when the Hubble constant falls below the mass of the axion, the axion rolls down its potential and begins to oscillate. At this point the dilaton is destabilized and also rolls down its potential. For $\lambda >1 $, this rolling can be \emph{fast-roll} with the energy density in $\phi$ predominantly in kinetic energy. The equation of state of $\phi$ in this phase is $w=1$ and the energy density redshifts as $a(t)^{-6}$.

This realizes early dark energy, wherein the onset of the decay of dark energy is due to the small axion mass, which is itself in agreement with general intuition from the string theory axiverse, and the fluid into which the early dark energy decays is simply the kinetic energy of the early dark energy field.

As a final note before we proceed, we remark that the motivation for this setup comes from a promising approach to cosmology in string theory. The existence of moduli fields is one of the few model-independent predictions of string theory (or more generally, extra-dimensional theories). The prevailing approach is to stabilize and subsequently ignore all these fields, in order to focus on the minimal fields necessary to describe our universe\footnote{Such as those that describe the position and excitations of D-branes, see e.g.~\cite{Alexander:2001ks,Burgess:2001fx,Dvali:2001fw} for early examples.}. However, the dynamics of these moduli fields can leave tell-tale signs on cosmological observables: In the context of inflation, this can lead to observable gravitational waves in small field inflation \cite{McDonough:2018xzh}. In the post-inflation universe, oscillations of the moduli fields can drive a period of matter domination \cite{Acharya:2009zt}. The present letter suggests that string moduli may also solve the Hubble tension.

The structure of this letter is as follows: In Section II we review the dynamics of axions and dilatons. In Section III we propose our mechanism for solving the Hubble tension, and in Section IV we show quantitatively the validity of this solution. We close with a discussion in Section V.

\section{The Cosmological Dynamics of Scalar Fields}

\label{FieldDynamics}

Axions were first posited as a solution to the strong CP-problem of the standard model \cite{Peccei:1977hh,Wilczek:1977pj,Weinberg:1977ma}. They were later realized to be excellent dark matter candidates \cite{Preskill:1982cy,Abbott:1982af,Dine:1982ah}, and later yet realized to be ubiquitous in theories of quantum gravity \cite{Svrcek:2006yi,Arvanitaki:2009fg,Cicoli:2012sz}. The basic premise is a scalar field with a continuous shift symmetry that is broken by non-perturbative effects to a discrete shift symmetry.

The potential for an axion field is given by
\be
V(\chi) = m_{\chi}^2 f^2 \cos \frac{\chi}{f} ,
\ee
and the equation of motion is,
\be
\ddot{\chi} + 3 H \dot{\chi} + V_{,\chi}=0 .
\ee
The canonical axion dynamics are as follows. Consider the axion initially displaced from the minimum of the potential. At early times, the Hubble friction dominates and the axion is slowly-rolling with $\dot{\chi}/\chi \propto (m^2/H) $. Once $H < m _\chi$, the Hubble friction term becomes sub-dominant and the axion rapidly rolls to the minimum and begins to oscillate.

Now we turn to the dilaton. In string theory the dilaton is related to the string coupling $g_s= e^{\phi}$, and combines with the fundamental axion to comprise the axio-dilaton field. For our purposes, the dilaton is interchangeable with the Kahler moduli fields, which parametrize the size of the extra dimensions, and when canonically normalized, appear in the four-dimensional potential via exponentials. More generally, the `dilaton' here can considered as simply a quintessence field, as widely studied in the dark energy literature (for a review see e.g.  \cite{Tsujikawa:2010sc,Tsujikawa:2013fta}). 

The dilaton potential is given by,
\be
V(\phi) = V_0 e^{- \lambda \phi/m_{pl}} . 
\ee
The parameter $\lambda$ is referred to as the ``slow-roll'' parameter \cite{Tsujikawa:2010sc}, $\lambda = m_{pl} |\partial_\phi V/V|$. The phase space dynamics of this model are well understood, and exhibits fixed points wherein the dilaton `tracks' the background. We will be interested in dynamics away from these tracking solutions.

 To understand the dynamics, let's again look for a slow-roll solution in a background where $\phi$ is subdominant and $H$ is a constant. If we assume the second derivative $\ddot{\phi}$ is initially sub-dominant, then the solution is given by,
\be
\label{phiSR}
\phi(t) = \frac{m_{pl}}{\lambda} \log \left[1 + \frac{\lambda ^2 t V_0}{3 H m_{pl} ^2}\right] .
\ee
To assess slow-roll, we can evaluate the sub-dominance of the second-derivative. At $t=0$:
\be
\frac{\ddot{\phi}}{H |\dot{\phi}|}|_{t=0} = \frac{V_0 \lambda^2}{3 H^2 m_{pl}^2} .
\ee
Thus when $V(\phi)$ is a subdominant contribution to $H$, i.e. $V_0 \ll H^2 m_{pl}^2$, the field will slow roll if $\lambda \lesssim1$.

However, if $\lambda \gg 1$, and with initial conditions away from a tracker fixed point, the solution is \emph{fast-roll}. During this fast-roll phase, the potential is subdominant  to the other terms in the equation of motion and the solution is simply
\be
\label{phiFR}
 \dot{\phi} \propto a(t)^{-3} .
\ee
Thus the kinetic energy redshifts as $a^{-6}$ and the energy initially stored in $V(\phi)$ is rapidly redshifted away. 

The fast-roll phase can not last forever, since $V_{,\phi\phi}\equiv m^2 _{\phi}$ is decreasing:
\be
\label{eq:endFR}
\frac{V_{,\phi \phi}}{H^2} = \frac{V_0 \lambda^2}{m_{pl}^2 H^2} e^{- \lambda \phi/m_{pl}} .
\ee
Thus, in a reversal of the dynamics of the axion, fast-roll will terminate when $V_{,\phi \phi}\sim H^2$, i.e. when Hubble becomes greater than the mass of the field. We must check then that the period of fast-roll can be sufficiently long-lived so as to dilute the early dark energy.

Setting equation \eqref{eq:endFR} equal to 1, this occurs when $\phi$ is given by:
\be
\phi_{end-FR} = \frac{m_{pl}}{\lambda} \log \left[ \frac{V_0 \lambda^2}{H^2 m_{pl}^2} \right] ,
\ee
at which point the energy density in the dilaton is given by,
\be
\rho_{\phi} \simeq V(\phi_{end-FR}) = \frac{H^2 m_{pl} ^2}{\lambda^2}.
\ee
And thus the dilaton will consitute a fraction the energy density of the universe given by $1/\lambda^2 \ll 1$. From this we conclude that the fast-roll phase can indeed dilute the energy density in the dilaton.

We can confirm these dynamics quantitivatively by solving numerically for the dynamics of $\phi$ in an FRW background. The equations of motion can be written as \cite{Tsujikawa:2013fta},
\bea
\frac{{\rm d}x}{{\rm d}N} && = - 3 x + \frac{\sqrt{6}}{2} \lambda y^2 + \frac{3}{2} x \left[ (1 - w_b)x^2 + (1+w_b)(1-y^2)\right] , \nonumber \\
\frac{{\rm d}y}{{\rm d}N} && = - \frac{\sqrt{6}}{2} \lambda x y + \frac{3}{2} y \left[ (1 - w_b)x^2 + (1+w_b)(1-y^2)\right] ,
\eea
with $x\equiv \dot{\phi}/(\sqrt{6} m_{pl} H)$, $y\equiv\sqrt{V(\phi)}/(\sqrt{3} m_{pl} H)$,  $N$ the number of e-folds, and $w_b$ is the equation of state of the background cosmology i.e.~the content of the universe aside from $\phi$).

We consider a matter-dominated universe $(w_b=0)$, and solve the system with $\lambda=400$ and initial conditions that the field is at rest and constitutes 10\% of the energy density of the universe, corresponding to $x(0)=0$ and $y(0)=1/\sqrt{10}$. In the context of a coupled axion-dilaton system, this applies as an approximate description once the axion has rolled down its potential. 

\begin{figure}[ht!]
\centering
\includegraphics[width=8.5cm]{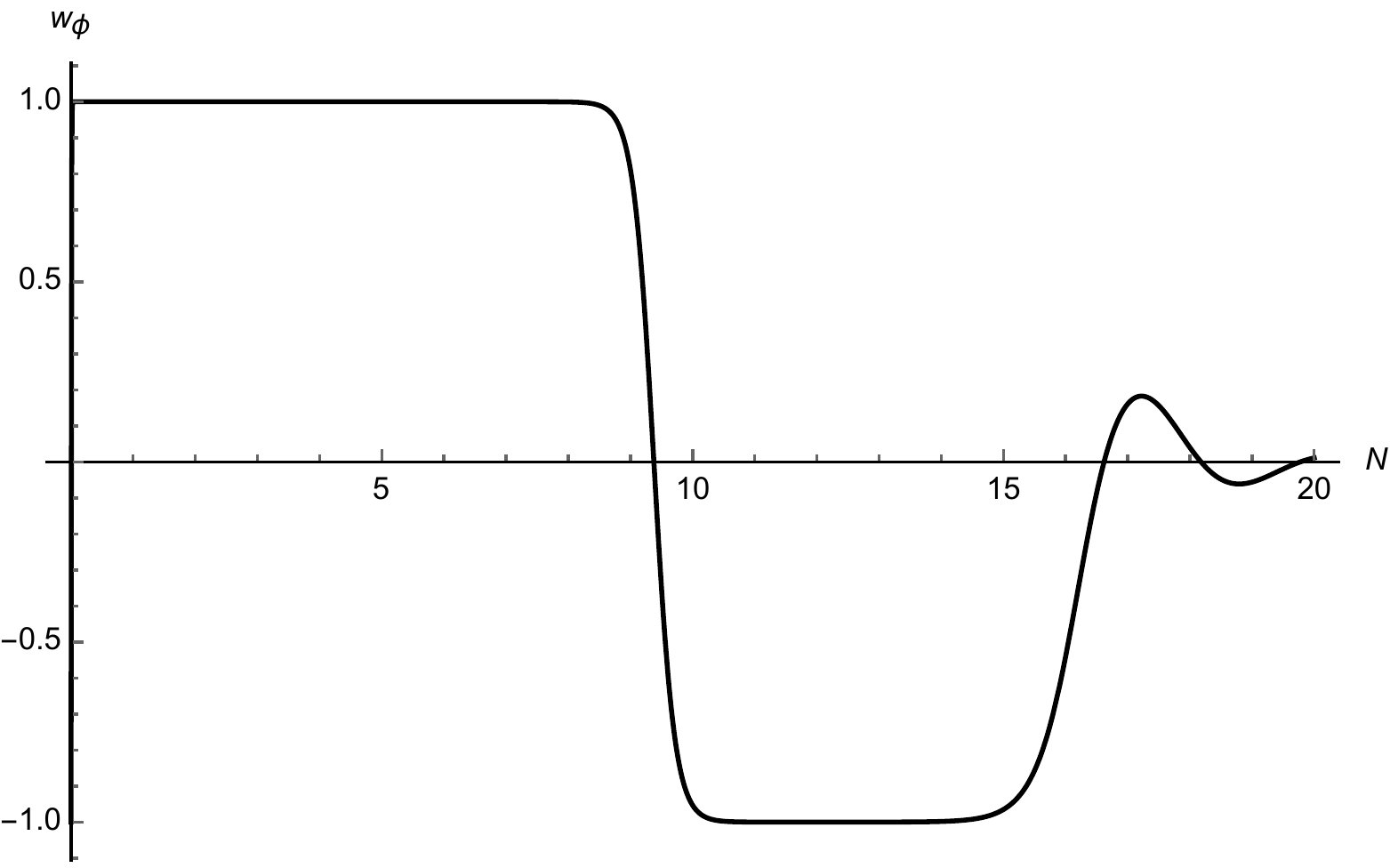}
\caption{Evolution of the equation of state for a quintessence field with potential $V=V_0 e^{- \lambda \phi/m_{pl}}$, with $\lambda=400$ and initial conditions $x(0)=0$, $y(0)=1/\sqrt{10}$, in a background with equation of state $w_b=0$. } 
\label{fig:wdilaton}
\end{figure}

\begin{figure}[ht!]
\centering
\includegraphics[width=8.5cm]{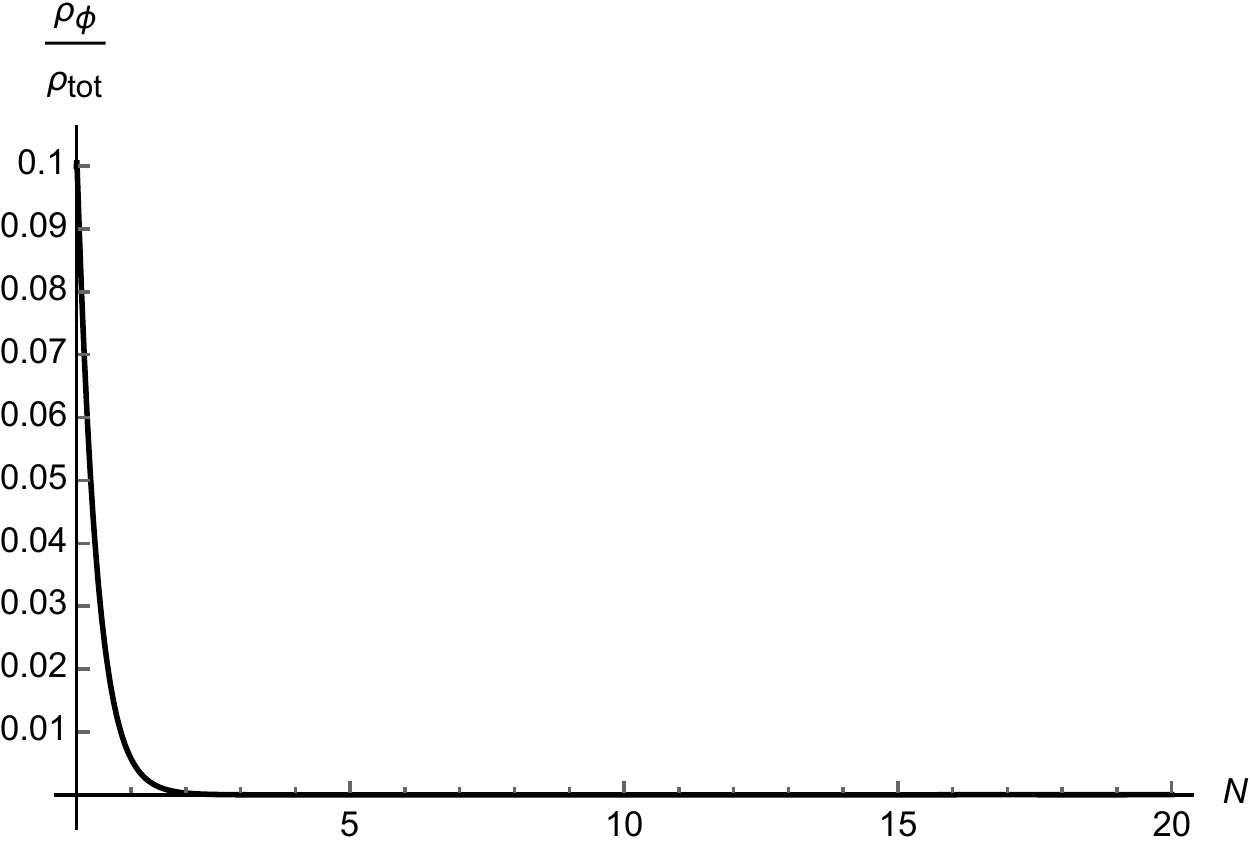}
\caption{Evolution of the energy density for a quintessence field with potential $V=V_0 e^{- \lambda \phi/m_{pl}}$, with $\lambda=400$ and initial conditions $x(0)=0$, $y(0)=1/\sqrt{10}$, as a fraction of the total energy density of the universe, in a background with equation of state $w_b=0$.} 
\label{fig:rhodilaton}
\end{figure}

In Figure \ref{fig:wdilaton} we plot the evolution of the equation of state, $w=(x^2 - y^2)/(x^2+y^2)$, which shows a rapid onset of fast-roll evolution ($w=1$), lasting for $\approx 9.5$ e-folds, followed by a slow-roll phase lasting $\approx 7$ e-folds, before finally landing on the tracker solution $w=w_b=0$. If $t=0$ is defined as being matter-radiation equality, $z_{eq}\simeq3000$, then the present time $z=0$ corresponds to $N = 8$ the present universe would be in the fast-roll phase. In Figure \ref{fig:rhodilaton} we plot the evolution of the energy density, as a fraction of the energy density of the universe, which shows the dilaton never does recover from the fast-roll dilution.

\vspace{.2cm}

\section{The Mechanism: Early Dark Energy from Axion-Dilaton Dynamics}

We now come to the coupled system. Consider the interacting axion-dilaton system,
\be
\label{V}
V(\phi,\chi) = \frac{1}{2}m_{\chi}^2 f^2 e^{\beta \phi/m_{pl}}  (1+\cos \frac{\chi}{f}) + V_0 e^{- \lambda \phi/m_{pl}} .
\ee
One could also include a coupling of the dilaton to the kinetic energy of the axion, but this effect is degenerate with the other two terms and so we ignore it. These couplings naturally arise between the axion and Kahler moduli of string theory, and the dilaton, related to the string coupling $g_s = e^{\phi}$ \cite{Cicoli:2012sz}, and similar Lagrangians have been discussed recently in the context of quintessence dark energy in string theory \cite{vandeBruck:2019vzd}. Axion interactions in dark energy models have also been studied in a different context in \cite{Alexander:2016mrw,Alexander:2016nrg}. In general there could also be couplings of $\phi$ to the standard model fields, which may have interesting effects, but which are highly model dependent, and can be suppressed relative to the coupling to dark matter, as has been discussed in the dark energy literature in  e.g. \cite{Gasperini:2001pc}.  For example, if the axion and $\phi$ are both bulk fields while the standard model is localized to a stack of intersecting branes, there can naturally be a hierarchy of the interaction strengths.

Importantly, the parameter $\beta$ is taken to be positive. This is in contrast with the usual intuition from string theory \cite{Dine:1985he}. To reconcile this, $\beta$ can be understood as a function of $\phi$, and other moduli fields, that is positive in a region of field-space near $\phi=0$, for example,
\be
\label{betaphi}
\beta \equiv \beta(\phi,\Phi^i) \simeq \beta  - \mathcal{O}(  \frac{\phi}{m_{pl}}) ,
\ee
where $\Phi^i$ are a set of other fields which determine the coefficients in the expansion of $\beta(\phi)$. We will see the phenomenological implications of this later.

The parameter $\lambda$ is taken to be greater then 1, to allow for the fast-roll dynamics observed in the previous section. This is in line with recent arguments from quantum gravity that $|V'|/V > \mathcal{O}(1)$ for all effective field theories that can be consistently coupled to gravity \cite{Obied:2018sgi}. More generally, quantum corrections to the $\phi$ potential will tend it \emph{more} steep, as per the standard $\eta$-problem \cite{Copeland:1994vg,Baumann:2014nda}, and thus will not alter the destabilization that is the basis of the mechanism in this paper.

We assume that the initial condition for the axion is that it is displaced from the minimum of its potential, as per the usual axion misalignment mechanism \cite{Preskill:1982cy,Abbott:1982af,Dine:1982ah}, and is frozen by Hubble friction, or rather, it is slowly-rolling. This determines the initial condition for $\phi$: The $e^\phi$ dependence of the axion potential stabilizes the dilaton at the minimum of the combined potential \eqref{V},
\be
\phi_0 = \frac{m_{pl}}{\beta + \lambda} \log \left[ \frac{V_0 \lambda}{\beta f^2 m_{\chi}^2 (\cos \chi/2f) ^2}\right] . 
\ee
For large $\lambda$, $\phi_0$ is very close to $0$. Thus the potential in the early universe, with the dilaton frozen at $\phi_0 \simeq 0$, is given by
\be
V_{early} = m_{\chi}^2 f^2 \left( 1+\cos \frac{\chi}{f} \right) + V_0 .
\ee
The constant $V_0$ is an additional contribution to the cosmological constant in the early universe.

Once the Hubble parameter becomes less than the mass of the axion, $H \lesssim m_{\chi}$, the axion rapidly rolls down its potential and begins to oscillate. The solution for the axion is given by,
\be
\chi(t) \simeq \frac{\chi_0}{a(t)^{3/2}} \cos m_\chi t  .
\ee
In order for this to occur shortly before recombination, we require $m_{\chi} \sim 10^{-27}$ {\rm eV}. Thus $\chi$ is an ultralight axion. It is constrained to be a sub-dominant component of the dark matter \cite{Hlozek:2014lca}.

\begin{figure}[ht!]
\centering
\includegraphics[width=8.5cm]{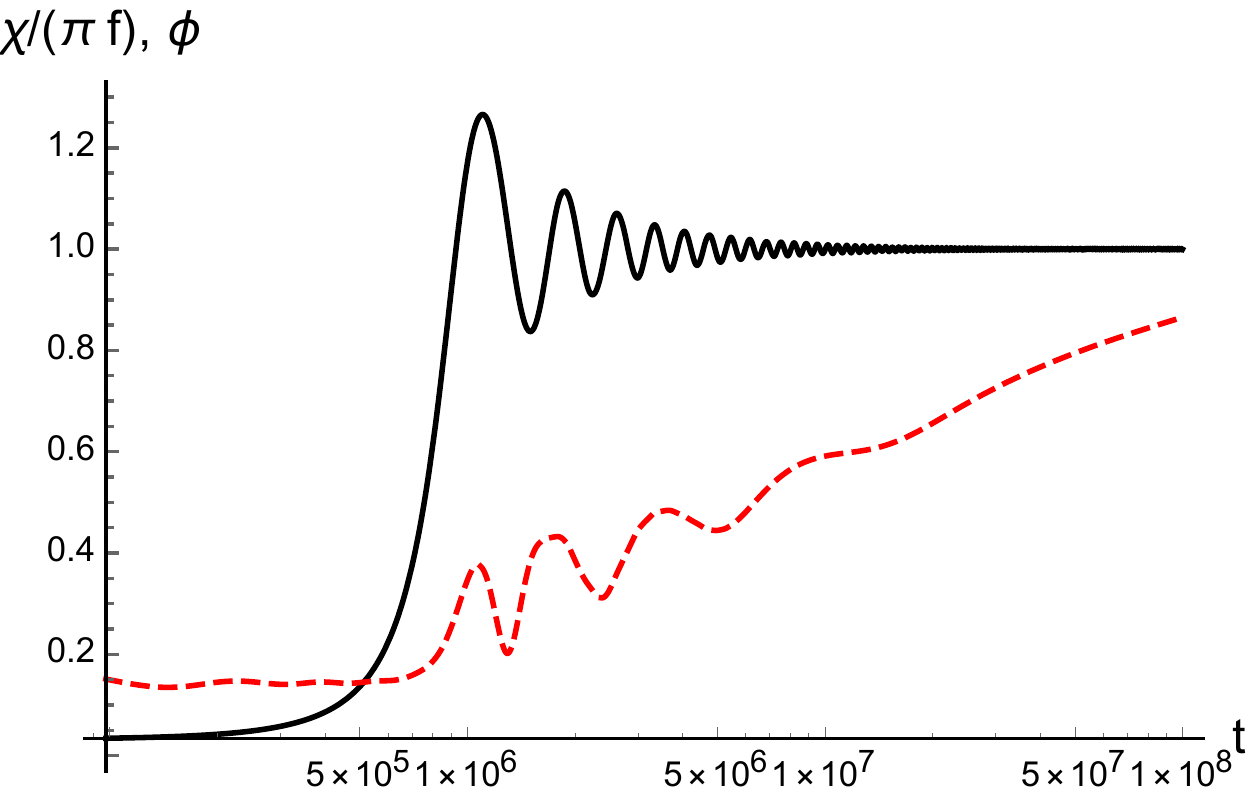}
\caption{Evolution of the fields in the coupled axion-dilaton system, with $\chi$ in units of $\pi f$ and $\phi$ in units of $m_{pl}$.  The axion (black, solid) begins to oscillate at $t \sim 10^6$ triggering a destabilization and subsequent fast-roll of the dilaton (red, dashed). } 
\label{fig:Fields}
\end{figure}

When $\chi$ drops down its potential, $\phi$ is destabilized and begins to roll. As this occurs, the effective axion mass ${m_{\chi} ^2 }_{eff} = m^2 _{\chi} e^{\beta \phi/m_{pl}}$ increases and $\chi$ undergoes extremely rapid damped oscillations.  In this phase $\chi$ is well approximated by its time-averaged value $\langle \chi \rangle = \pi/f$. The dynamics of $\phi$ are then well described simply by the dilaton potential,
\be
V_{\phi} =  V_0 e^{- \lambda \phi/m_{pl}} .
\ee
As shown in section \ref{FieldDynamics}, for $\lambda \gg 1$ the subsequent dynamics of $\phi$ are kinetic energy dominated and hence the energy evolves density as
\be
\rho_{\phi} \simeq \frac{V_0}{a^6} ,
\ee
and thus quickly redshifts away.

\begin{figure}[ht!]
\centering
\includegraphics[width=8.5cm]{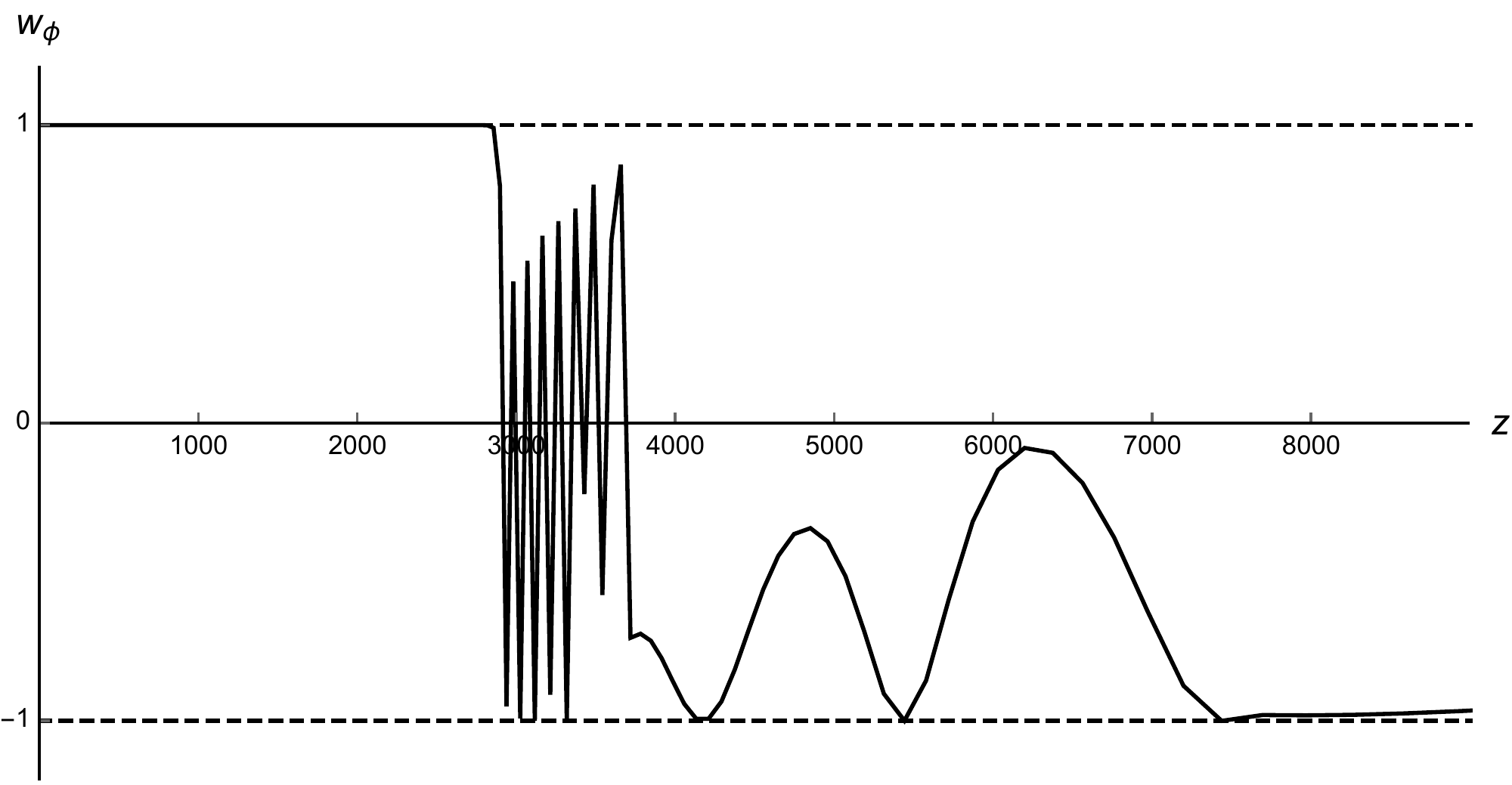}
\caption{ Redshift evolution of the dilaton equation of state in the coupled axion-dilaton system \eqref{V}, in a matter dominated universe. Time flows right to left, with the present time given by $z=0$. Dashed lines are $w=-1$ and $w=+1$, corresponding to slow and fast roll respectively.} 
\label{fig:w}
\end{figure}

To demonstrate these dynamics, we numerically solve this system in a matter-dominated universe (for simplicity neglecting radiation, keeping only matter and the axion-dilaton system) for a set of parameters, in natural units.  We initialize the system at  $t=t_*$ with $H(t_*)=H_*=10$, $m_\chi/H_{*} = 10^{-6}$, $\chi_*=f/10$, $f/H_*=0.08$,  $\phi_*=0.2 m_{pl}$ and $V_0 = m_{\chi}^2 f^2$. We fix the exponents as $\beta=1$ and $\lambda=20$, which control the damping of the oscillations and the speed of the fast-roll evolution respectively.  The result is shown in Figure \ref{fig:Fields}, where we observe the destabilization of the two fields, and the onset of fast roll evolution of the dilaton.

To better understand the transition and ensuing evolution, we plot the dilaton equation of state in Figure \ref{fig:w}, where we use redshift $z$ as the dependent variable, with $z=0$ (today) defined such that the transition occurs around $z=3000$. Figure \ref{fig:w} shows a transition from $w=-1$ to a long-lived phase of $w=+1$, and thus a change in the dynamics of $\phi$ from dark energy to kinetic energy.  These two regimes are separated by a transient phase of oscillations, whose total time duration, as well as the frequency and amplitude of oscillations, depends sensitively on the parameters of the axion potential. Such oscillations are familiar from Early Dark Energy \cite{Karwal:2016vyq,Poulin:2018cxd}, and are irrelevant for observation insofar as they are rapid enough to be averaged over (see also \cite{Agrawal:2019lmo} for a discussion).

One can also understand how the dynamics are changed by modifications to the coupling $\beta$, for example $\beta \rightarrow \beta(\phi)$ as in \eqref{betaphi}. We repeat the previous numerical example, but modify $\beta$ as
\be
\label{betaphic}
\beta(\phi) = \beta - c \frac{\phi}{m_{pl}} .
\ee
In this case, the $\chi$-induced minimum for $\phi$, which only exists for $\beta>0$, disappears once $\phi > (c/\beta) m_{pl}$.

\begin{figure}[ht!]
\centering
\includegraphics[width=8.5cm]{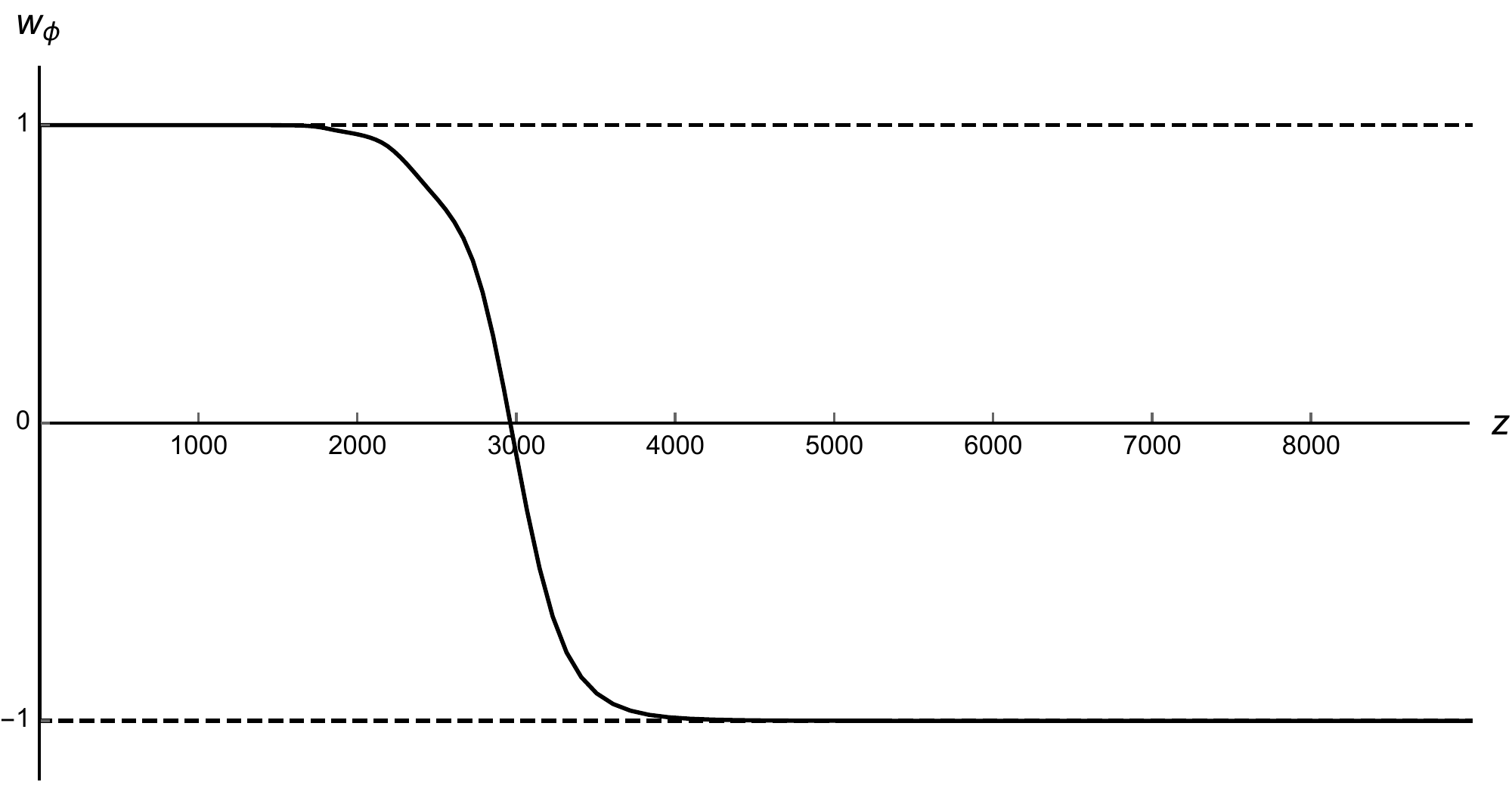}
\caption{ Redshift evolution of the dilaton equation of state for the modified axion-dilaton coupling \eqref{betaphic}, in a matter dominated universe. Time flows right to left, with the present time given by $z=0$. Dashed lines are $w=-1$ and $w=+1$, corresponding to slow and fast roll respectively.} 
\label{fig:wmodelc}
\end{figure}

The resulting equation of state, for  $\beta=c=1$ and $\lambda =15$, is shown in Figure \ref{fig:wmodelc}. The oscillations in $w$ are no longer present, and the system exhibits a rapid yet smooth transition from slow- to fast-roll.

 \vspace{-.4cm}

\section{Resolving the Hubble Tension}

We now turn to the main goal of this letter. To realize the background evolution of the Early Dark Energy resolution of the Hubble tension, we require that the early dark energy $V_0 e^{- \lambda \phi_0 /m_{pl}}\simeq {\rm eV}^4$ be rapidly dissipated around recombination, $z_{rec}\sim 10^3$. The latter requirement sets the axion mass $m_{\chi} \lesssim H_{rec}$, which is given by $H_{rec}  \simeq {\rm eV}^2/ m_{pl} \sim 10^{-27} \rm{eV}$. 

The former requirement then relates the normalization $V_0$, the decay constant $f$, and the exponents $\lambda$, $\beta$. We assume for simplicity that $\lambda \gg \beta$, in which case the condition $V_0 e^{- \lambda \phi_0 /m_{pl}}\simeq {\rm eV}^4$, using the explicit expression for $\phi_0$, reads
\be
f \simeq \sqrt{\frac{\lambda}{\beta}} \frac{{\rm eV}^2}{m _{\chi}} ,
\ee
with $V_0$ left as a free parameter. For $m_{\chi}\sim 10^{-27} {\rm eV}$, $f$ is of order the Planck scale. Curiously, that a Planckian field excursion should cause a destabilization is precisely the content of an another conjecture based in quantum gravity: the swampland distance conjecture \cite{Ooguri:2006in}.

A complete analysis of the model requires a numerical evolution of perturbations. This was studied in great detail in \cite{Agrawal:2019lmo}, where it was shown that it does not suffice to treat an oscillating early dark energy field as an effective fluid, but rather a proper accounting of perturbations must solve the perturbed field equations. We leave an analogous analysis for the present model for future work.

{\bf Comment added after publication:} After publication of this work, an analysis of perturbations by \cite{Smith:2019ihp} revealed that the effective fluid description does in fact capture the main features of the EDE scenario.  See \cite{Smith:2019ihp} for more details.
 
\section{Discussion}

In this letter we have proposed that the dynamics of a coupled axion-dilaton system can resolve the Hubble tension by providing a realization of Early Dark Energy. In this scenario the canonical evolution of the axion leads to a destabilization of the dilaton, inducing a phase of fast-roll in the latter, and the conversion of potential energy into kinetic energy, which subsequently redshifts away as $a^{-6}$. For an axion of mass $10^{-27} {\rm eV}$, this conversion occurs around recombination, similar to the dynamics of \cite{Agrawal:2019lmo}. 

Our work proposes a solution to the Hubble tension based in standard particle cosmology. In an upcoming work we will provide a detailed numerical computation of the cosmological evolution, and a complete perturbation analysis including a Markov-Chain Monte-Carlo analysis of the model, as well as an exploration of the parameter dependence and phase-space dynamics, and the impact of this on the transient phase of oscillations that separates the early dark energy and fast-roll phases.  We also plan to study of the genericity of this mechanism and parameters in string theory and supergravity \cite{followup}.



\acknowledgments
 \vspace{-.1cm}
We would like to thank Alexandre Barreira, Robert Brandenberger, Yifu Cai, Elisa Ferriera, Jim Gates, Alan Guth, Eiichiro Komatsu, Christina Kreisch, Antonino Marciano, David Pinner, Lisa Randall, Cumrun Vafa, David Spergel, Kyriakos Vattis, and Matias Zaldarriaga, for helpful discussions.  We also thank a referee for many useful comments and suggestions. EM is supported in part by the National Science and Engineering Research Council of Canada via a PDF fellowship.  

\bibliography{axion-dilaton-refs}
\bibliographystyle{JHEP}

\end{document}